\documentclass{article}

\usepackage{arxiv}

\usepackage[utf8]{inputenc} % allow utf-8 input
\usepackage[T1]{fontenc}    % use 8-bit T1 fonts
\usepackage{hyperref}       % hyperlinks
\usepackage{url}            % simple URL typesetting
\usepackage{booktabs}       % professional-quality tables
\usepackage{amsfonts}       % blackboard math symbols
\usepackage{nicefrac}       % compact symbols for 1/2, etc.
\usepackage{microtype}      % microtypography
\usepackage{lipsum}		% Can be removed after putting your text content
\usepackage{graphicx}
\usepackage{natbib}
\usepackage{doi}

\title{Voice CMS: updating the knowledge base \\of a digital assistant through conversation}

\author{ \href{https://orcid.org/0009-0006-4590-9236}{\includegraphics[scale=0.06]{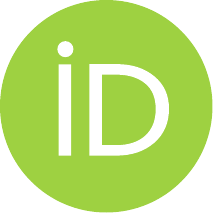}\hspace{1mm}Grzegorz~Wolny} \\
	Orange Research\\
	Warsaw, Poland \\
	\texttt{grzegorz.wolny@orange.com} \\
	\And
	\href{https://orcid.org/0009-0000-6323-4806}{\includegraphics[scale=0.06]{orcid.pdf}\hspace{1mm}Michał K.~Szczerbak} \\
	Orange Research\\
	Warsaw, Poland \\
	\texttt{michal.szczerbak@orange.com} \\
}

\renewcommand{\shorttitle}{Voice CMS - G. Wolny and M. Szczerbak}

\hypersetup{
pdftitle={Voice CMS: updating the knowledge base of a digital assistant through conversation},
pdfsubject={cs.HC},
pdfauthor={Grzegorz~Wolny, Michał K.~Szczerbak},
pdfkeywords={voice user interface, knowledge management, digital assistant, bot, user experience, human-ai interaction, knowledge-grounded conversation, knowledge extraction, comparative / qualitative study},
}

\begin{document}
\maketitle

\begin{abstract}
	In this study, we propose a solution based on a multi-agent LLM architecture and a voice user interface (VUI) designed to update the knowledge base of a digital assistant. Its usability is evaluated in comparison to a more traditional graphical content management system (CMS), with a focus on understanding the relationship between user preferences and the complexity of the information being provided. The findings demonstrate that, while the overall usability of the VUI is rated lower than the graphical interface, it is already preferred by users for less complex tasks. Furthermore, the quality of content entered through the VUI is comparable to that achieved with the graphical interface, even for highly complex tasks. Obtained qualitative results suggest that a hybrid interface combining the strengths of both approaches could address the key challenges identified during the experiment, such as reducing cognitive load through graphical feedback while maintaining the intuitive nature of voice-based interactions. This work highlights the potential of conversational interfaces as a viable and effective method for knowledge management in specific business contexts.
\end{abstract}

% keywords can be removed

\section{Motivation}
Last couple of years have revolutionized the field of bots thanks to the rapid advancement on the topic of large language models (LLMs), the main promise being more natural conversations between the assistants and humans \cite{koubaa:gpt}. In fact, the LLM technology is replacing more and more classical bot-based products based on intent recognition models, which required more curated knowledge bases using conversation trees and often resulted in very inflexible dialogues that users needed to adjust to. User experience is thus designed to improve with the bots employing LLMs.

The opposite happens, however, when one detects that behind well-rounded words providing a perfect match to the questions asked hides a lie \cite{kim:dissat} \cite{oelschlager:trust}. It is only natural for the LLMs, due to their very nature and the way they were trained, that they invent a chain of words not related to any truth if they don't have access to the true information. To cope with the problem of not having particular knowledge at the time of training an LLM, different methods were introduced, including research augmented generation (RAG), critical reasoning, tools, etc. \cite{tonmoy:halu}. Still, the quality of the information provided to LLM-based conversation systems plays a major role in creating trustworthy assistants.

What if this knowledge is not fixed, though? What if it changes slightly but constantly at a pace of company decisions and the ever-changing business context? We consider a videobot in the role of a customer-facing digital assistant used in a hotel lobby to provide the guests with up-to-date information about the hotel events or its restaurant, for example. Whereas the city-level attractions calendar can be provided by the specialized service provider, through an API, for instance, assuring that the digital assistant's knowledge is always valid falls into the responsibility of the hotel staff, who has the information in the first place.

Indeed, keeping a bot degenerating in time with obsolete information and misleading users can very swiftly become more of a problem than a benefit to any company. On the other hand, we need to consider that it might be too much to ask from the hotel employees to assure that the digital assistant's knowledge is always valid, should the method of doing so be too complex. Our current research focuses, therefore, on providing users with an effective interface that would facilitate the task of maintaining the knowledge base of digital assistants in the service of a real small to medium-sized company. We take an example of a hotel in Poland, providing us with an environment for our tests and experiments on digital assistants.

In this particular work, we explore employing an interface that is natural for both humans and digital assistants - namely, the voice user interface (VUI). In fact, it is already the interface used between the end users seeking information and our videobot, which we describe in more detail in \cite{lesiak:digiass}. Having surveyed the owners of the hotel it seems to them that adding a new graphical interface, in a form of a dedicated web application for example, would be ineffective for reasons of needing to first learn it, then log into it every time a new event needs to be added or menu changes in the restaurant. They assume that passing information vocally, like one would do to a junior co-worker, would augment the chances of employees actually transmitting the information to the system.

In this paper, we describe a solution to the challenge of creating both a backend system for managing the incremental knowledge base for the LLM-based digital assistant and a frontend VUI augmented with graphical elements to enhance the experience, which would answer the business need of the hotel. Moreover, we report on the experiment aiming to study the conditions under which the vocal interface can prove useful as complementary to the classical GUI-based content management systems (CMS). Ultimately, we aim to answer the research question (RQ): what is the difference in user experience between GUI and VUI in the context of providing new content into a system, and what does it depend on?

The remainder of this document is structured as follows. First, we give a relevant and recent literature review in Section~2. Then, in Section~3, the proposed solution is described. Section~4 outlines the design from the hypothesis, through setup to the actual execution of the experiment, the results of which are presented in Section~5. This leads us to the discussion of both the contributions of our findings and the limitations in Section~6. Finally, we conclude with Section~7 by distilling the main messages and suggesting future research dimensions.

\section{Background}
Management of the knowledge base for a digital assistant is still more often than not restricted to its off-line preparation, where a assistant is designed either to have some fixed and infrequently changing information base or to be linked to an external data source, i.e. in RAG solutions for LLMs. Hence, any even slight changes to the knowledge require intervention on the data source and most probably at least some proficiency in development or digital data manipulation. 

We have identified only one work from \cite{grassi:care} focusing on adding new knowledge into a conversation agent at run-time. Their use case was focused on a humanoid robot in a care home and was technologically based on a pre-LLM knowledge base grounded in an ontology. The authors report on the experiments aiming at the technical performance assessment concerning new concepts extraction and four methods developed to insert the recognized concepts into the ontology. They did not include, however, usability tests of teaching the system new topics via voice.

Another case study of inserting data into the system through a VUI is discussed in \cite{olivares:medi}. This method is introduced as an alternative to a traditional GUI in the context of a medical visit, where doctors should be focused on the patients and the consultation itself instead of the screen for capturing the information. There is a strong parallel here with our hotel use case, even if the latter concerns rather operational efficiency and aims to impact the knowledge base of the voice assistant itself.

Another graphical interface extension with a vocal one is studied in \cite{reicherts:colab}, where collaborative tasks for interpreting different visualizations on a screen were performed with or without the use of a vocal assistant, effectively giving insights for differences in interactions that humans engage with GUIs vs VUIs. Ultimately, the results show the effectiveness of voice interfaces in the context of human-like collaboration. This is an encouraging conclusion in the case of an assistant employed in a hotel, which in the context of updating the knowledge base might be treated by the staff as a junior co-worker.

There have been several other studies in the last two decades looking into evaluating voice interfaces, frequently in comparison to graphical interfaces. Both \cite{chavez:siri} and \cite{zhang:iphone} report on preliminary experiments around several tasks performed on iPhones and promising results for VUI in terms of effectiveness and user preference. Even more optimistic results are discussed in \cite{chandel:literate}, where the voice part of the system was prepared in a Wizard of Oz fashion and focused on the differences between more or less literate users. An earlier study showed, however, that even though a vocal interface might be preferred and work faster, it may also be less accurate and more error-prone \cite{damianos:msiia}.

It should be noted that some of the previously mentioned publications refer to relatively old first attempts to compare the two types of interfaces on a very different maturity level at that time. A more recent study from the authors of \cite{cha:context} emphasizes the importance of activity context on the preference for choosing voice or touch. The reference GUI in that experiment consisted of buttons with Korean text, and researchers were observing a user preference switching point depending on the number of syllables per button, which seems to be a very specific use case.

On the other hand, \cite{limerick:agency} gives a little more general perspective on the problem of employing VUIs, as they are proven to suffer from a lesser sense of agency when compared to GUIs. The authors argue that this does not depend on the voice-related technology and should hold true despite its progress. This relates to the human natural preference for voice for cooperation and graphical interactions for focusing on the task \cite{lebigot:coop}, and the question asked by the researchers in \cite{qvarfordt:tool} regarding whether the interfaces should become human-like or stay tool-like. In the latter study, we learn that the more anthropomorphic the voice feedback was given to experiment participants, the more they preferred the system to be human-like.

In our work, we wish to further deepen the understanding of human preferences regarding VUIs and the factors that impact both expressed satisfaction and measured cognitive load for such complex communication tasks as updating a digital voice assistant's knowledge base in a very specific business setting.

As voice interfaces are gaining more and more interest, especially in the context of modern digital assistants using LLM-based conversational systems, there is growing attention to better evaluate their usability \cite{deshmukh:survey}. Some researchers focus on deriving new measures and qualities specific to digital virtual assistants, going away from the focus on effectiveness \cite{dutsinma:iso} and towards correlation with user satisfaction, intention for continuous use and recommendation \cite{chen:dvaux}. Others advocate for preparing new VUI design guidelines to address the differences from well-studied GUIs \cite{nowacki:ergo} \cite{murad:guide}, which we implement in our solution.

\section{Solution}

\begin{figure}[htbp]
    \centering
    \begin{minipage}[b]{0.32\textwidth}
        \centering
        \includegraphics[width=\textwidth]{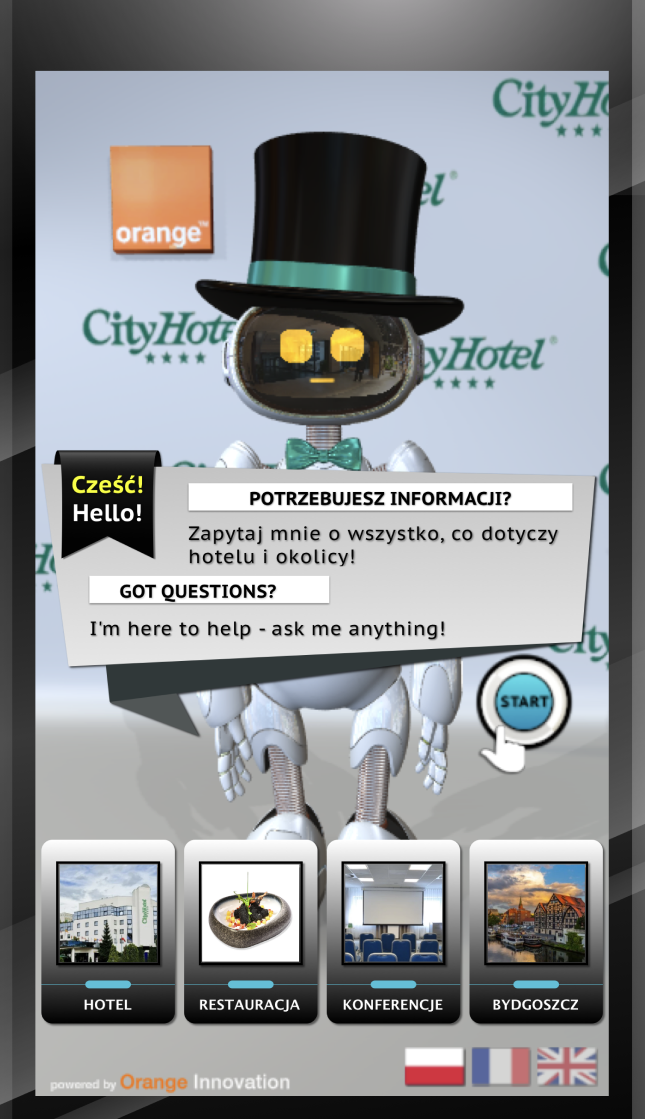}
    \end{minipage}
    \hfill
    \begin{minipage}[b]{0.32\textwidth}
        \centering
        \includegraphics[width=\textwidth]{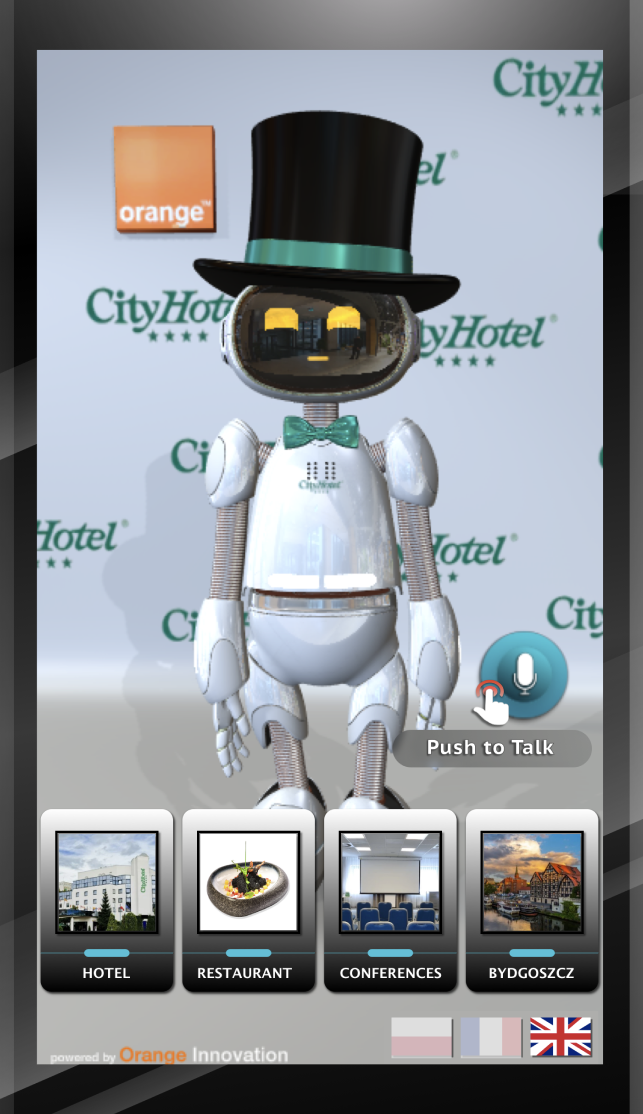}
    \end{minipage}
    \hfill
    \begin{minipage}[b]{0.3\textwidth}
        \centering
        \includegraphics[width=\textwidth]{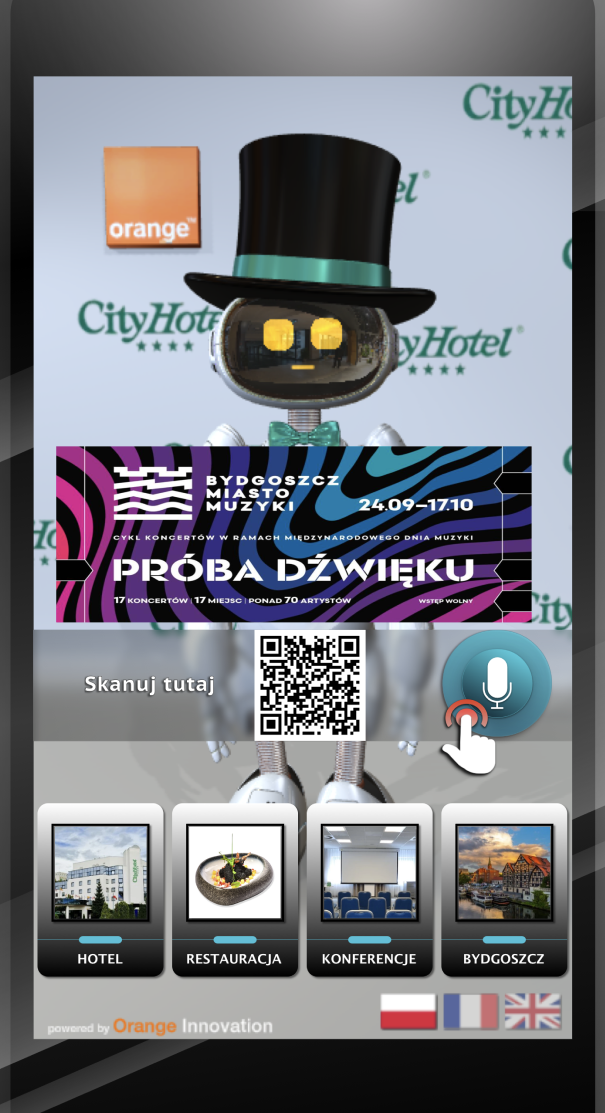}
    \end{minipage}
    \caption{Digital assistant}
    \label{fig:digital_assistant}
\end{figure}

The solution presented in this study builds upon a digital assistant equipped with a voice user interface (VUI). The assistant, implemented as a 3D-rendered digital character displayed on a large touchscreen in a hotel lobby (see Figure~\ref{fig:digital_assistant}), serves as a customer-facing videobot. Its primary role is to provide hotel guests with up-to-date information about the hotel’s offerings, assist in selecting menu items from the hotel restaurant, recommend nearby attractions, and help plan routes by displaying QR codes linking to navigation tools. The assistant is developed using the Unity engine, which not only renders the 3D character in real time but also supports the display of graphical elements such as banners, buttons, and QR codes. The system integrates speech recognition and synthesis capabilities, while the conversational engine is built on a multi-agent architecture using LangGraph\footnote{https://www.langchain.com/langgraph}. This architecture employs nodes to define workflows aimed at responding to user queries, with most of the nodes using large language models, specifically Gemini~1.5~Flash\footnote{https://gemini.google.com/}, for natural language interpretation and generation.

To ensure the assistant remains a reliable source of information, a method for providing and maintaining its knowledge base is required. This posed a challenge, as introducing a new IT system for knowledge management would impose a new obligation on hotel staff, requiring additional training and time investment. To address this, a voice-based content management system (Voice CMS) was developed, allowing staff to update the assistant’s knowledge base through natural conversation. This approach aligns with the intuitive interaction style already employed by the assistant when communicating with guests, thus reducing the cognitive and operational load on hotel employees.

\begin{figure}
    \centering
    \includegraphics[width=0.65\linewidth]{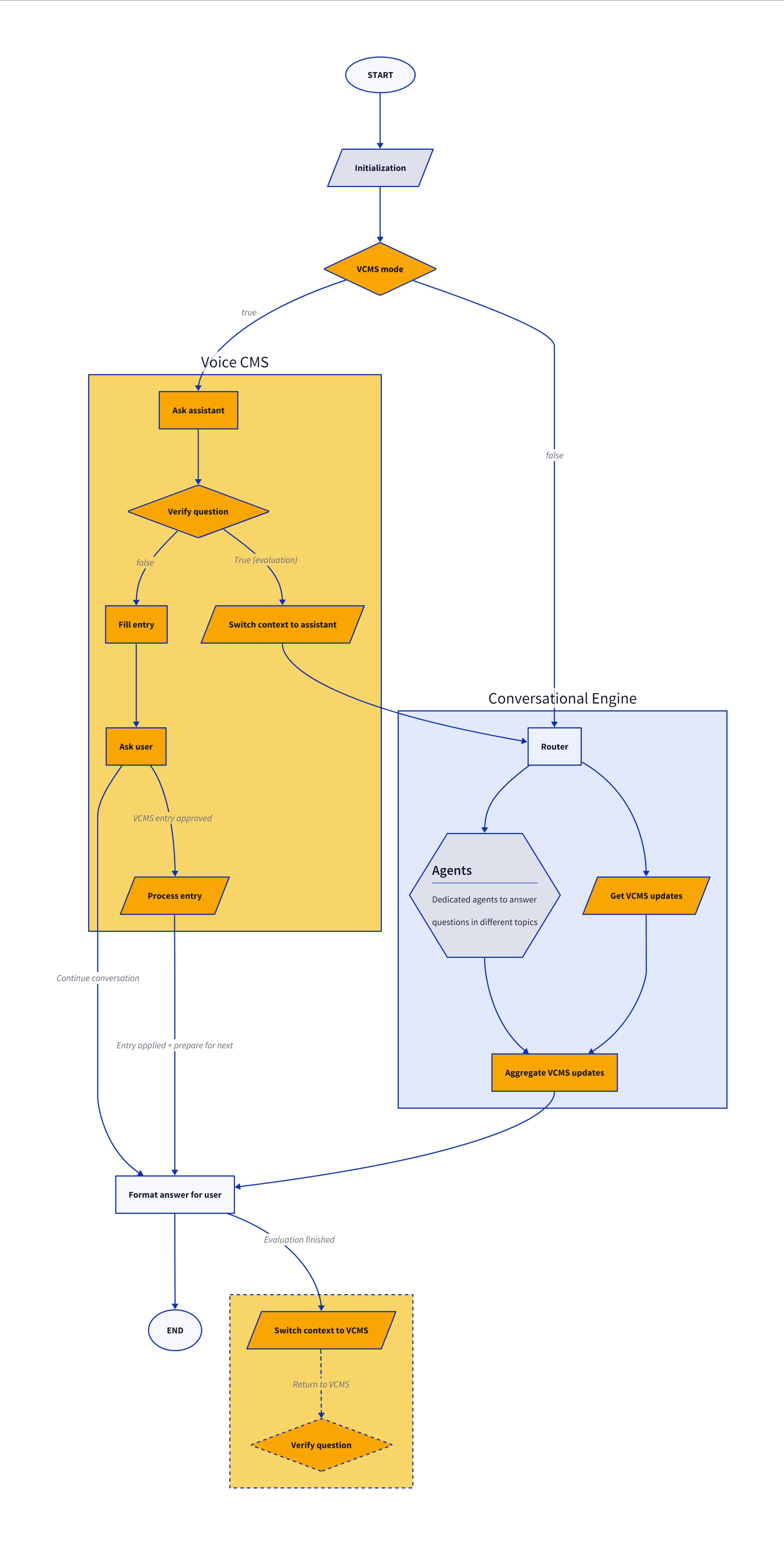}
    \caption{Voice CMS architecture}
    \label{fig:vcms_architecture}
\end{figure}

The Voice CMS operates in a secure mode, activated by tapping a specific area of the touchscreen and entering an access code. The solution's architecture is illustrated in Figure~\ref{fig:vcms_architecture}, where the nodes represented as rectangles leverage large language models to operate. Once activated (\texttt{true} branch after \texttt{VCMS mode}), hotel staff can add new information by conversing with the assistant. During the initial phase of the interaction, the assistant summarizes the information provided, ensuring clarity and completeness. If the assistant detects some details are unclear or missing, it can actively prompt the user for clarification (node \texttt{Ask user}). Simultaneously, the system constructs a draft entry for the VCMS in the background, which includes details such as the validity period of the information, its category, and the timeframe during which the assistant can use it in guest interactions (node \texttt{Fill entry}). Once the information is complete, the assistant vocally presents a full summary for confirmation. At this stage, staff can make corrections or approve the entry. Upon confirmation, the information is added to the knowledge base in the form of a JSON object, and the assistant is ready to receive another piece of information (node \texttt{Process entry}). It is worth noting that the Voice CMS mode can query the existing conversational engine to supplement knowledge or build context for information currently being processed in the conversation (node \texttt{Ask assistant}).

The Voice CMS is designed as an independent module integrated with the previously existing conversational engine. This modular approach ensures compatibility with other conversational systems, as the knowledge updates from the Voice CMS are aggregated with the assistant’s existing knowledge base at a single integration node in the LangGraph workflow (node \texttt{Aggregate VCMS updates}). This design enhances the system’s flexibility and scalability, making it suitable for deployment in various contexts beyond the hotel industry. 

By enabling natural, conversational updates to the knowledge base, the Voice CMS addresses the dual challenge of maintaining the assistant’s reliability while minimizing the operational burden on hotel staff. This solution not only ensures that the assistant remains a trustworthy source of information but also aligns with the intuitive interaction style.

\section{Evaluation}
This study investigates the usability and effectiveness of voice and graphical interfaces in completing tasks of varying complexity, with a focus on their application in knowledge management systems for digital assistants. The primary objective is to evaluate user preferences and performance when interacting with these interfaces, particularly in scenarios requiring structured information input, such as date handling and content categorization. While the proposed solution utilizes a voice interface for natural conversational updates, a GUI was also developed to perform a comparative study between the two approaches. Both interfaces are designed to serve the same purpose: facilitating the efficient and accurate maintenance of the digital assistant’s knowledge base.

%\paragraph{Research Hypothesis and Assumptions}
The central hypothesis posits that as task complexity increases, users will prefer the voice interface over the graphical interface. This assumption is based on the premise that voice interfaces allow for natural, conversational input, reducing the cognitive effort~\cite{schmidhuber:cognitive} required to structure and enter complex information. The voice system further enhances usability by interpreting input in real-time, extracting key details, verifying completeness, and prompting for clarification when necessary. In contrast, following the logic behind our hypothesis, graphical interfaces, while effective for simpler tasks due to their visual feedback and perceived control, may become cumbersome for complex tasks requiring lengthy or complicated inputs.

%\paragraph{Task Design and Complexity}
The experiment involved nine tasks, categorized into three arbitrary levels of complexity: easy, medium, and complex. Task complexity was determined by two factors: the length of the information to be entered and the level of detail of the content. Easy tasks required short, straightforward inputs, such as reporting a single event or issue. Medium tasks involved more detailed information, including specific dates and descriptions. Complex tasks required lengthy inputs combining multiple aspects, such as price changes, item removals, and the introduction of new items with detailed descriptions. Each task was designed to reflect real-world scenarios.

For example, an easy task involved reporting a jacuzzi malfunction valid for the current day. A medium task required announcing a beekeepers' fair with specific dates, times, and details about the event. A complex task involved updating a restaurant menu, including price changes, the removal of an item, and the introduction of a new dish with ingredients and pricing. These tasks were designed to test the system’s ability to handle varying levels of complexity and the users’ ability to interact effectively with each interface.

To approximate a neutral starting point, participants were given equal exposure to both interfaces through detailed instructions and a warm-up task to familiarize themselves with the systems. The warm-up task, which included example solutions for both interfaces, was excluded from the final analysis but served to reduce the learning curve and ensure participants were equally prepared as far as the understanding of both interfaces is concerned.

%\paragraph{Experimental Procedure}
The experiment was conducted using a web-based platform that ensured consistency across all participants. The platform randomized the order in which participants interacted with both interfaces and ensured that all required data were collected. Participants completed the same nine tasks in a fixed order ($[M, M, E, C, C, E, C, M, E]$, where the letters translates into \textbf{E}asy, \textbf{M}edium, \textbf{C}omplex), which was pre-randomized with respect to task difficulty to minimize learning effects. This controlled setup allowed for a reliable comparison between the two interfaces, isolating the effects of task complexity and interface design on user performance and preferences.

%\paragraph{Data Collection and Metrics}
After each task, participants rated its difficulty using the Single Ease Question (SEQ) \cite{sauro:seq}. Upon completing all tasks for a given interface, participants evaluated its overall usability using the System Usability Scale (SUS) \cite{brooke:sus}. Finally, after completing both sets of tasks, participants indicated their preferred interface (GUI, Voice CMS, or no preference) for each of the tasks and were also given the opportunity to share any additional observations or feedback in an open-ended comment field. In addition to these subjective measures, the study also collected objective metrics to assess interface performance. These included task completion time and the quality of task execution, which was rated on a 5-point scale based on predefined rules (handled manually in post-processing). Ratings of 4 or 5 indicated that the information produced was suitable for guest-facing dialogues, while lower scores reflected varying degrees of inaccuracy or incompleteness, with a score of 1 indicating that the content was essentially unusable. 

For the Voice CMS, further interaction-specific data were gathered, such as the number of user utterances during the dialogue, the need for corrections after presenting summaries, and whether the summaries were utilized. These supplementary metrics provided deeper insights into user interaction with the voice interface.

\section{Results} 

This section presents the analysis of data collected from a user study comparing GUI and Voice CMS. The study involved 7 participants (2 female, 5 male), aged between 33 and 51 years (M = 42, SD = 6.1). Each participant completed a series of tasks using both interfaces, along with corresponding usability surveys. All participants successfully finished the study, yielding a complete and balanced dataset of 126 task observations (63 for GUI, 63 for Voice CMS).

\subsection{Descriptive Statistics}

Overall usability assessments indicated a preference for the graphical interface. The mean SUS score for the GUI was 78.6 (SD = 21.9, range [40, 100], N=7), falling in the 'Good' range according to usability benchmarks~\cite{bangor:sus_adjective} and approaching 'Excellent'. The Voice CMS received a mean SUS score of 67.5 (SD = 22.1, range [32.5, 95], N=7), which is close to the historical average for usability studies~\cite{sauro:book}, often interpreted as 'Okay' or marginally 'Good'.

Examining task-level metrics, objective performance (\texttt{score}, scale 1-5) was generally high for both interfaces, though slightly better for the GUI (M = 4.54, SD = 0.74) than for the Voice CMS (M = 4.40, SD = 0.89). A more distinct difference emerged in subjective ease-of-use (\texttt{seq\_score}, scale 1-7, where 7=Very Easy). Participants rated tasks performed with the GUI as noticeably easier (M = 5.86, SD = 1.27) compared to the Voice CMS (M = 5.10, SD = 1.73). As expected, perceived ease-of-use decreased for both interfaces as nominal task difficulty increased; for instance, mean SEQ scores dropped from 6.52 (GUI) and 6.00 (VCMS) for 'Easy' tasks to 5.05 (GUI) and 4.14 (VCMS) for 'Complex' tasks. Efficiency metrics also favoured the GUI, which yielded shorter average processing times (M = 179 s, SD = 124 s) compared to the Voice CMS (M = 204 s, SD = 134 s). Task completion time predictably increased with task difficulty for both systems.

These initial descriptive results suggest potential advantages for the GUI in terms of overall usability, perceived ease-of-use, and efficiency, although objective performance was comparable. The following sections delve deeper into these findings through inferential statistical modeling, exploring how factors such as task difficulty, subjective perceptions, and task completion time influence user interface preference, while accounting for the dependencies inherent in the repeated-measures study design.

\subsection{Influence of Task Difficulty on Preference}

\begin{figure}[thbp]
   \centering
   \includegraphics[width=0.6\linewidth]{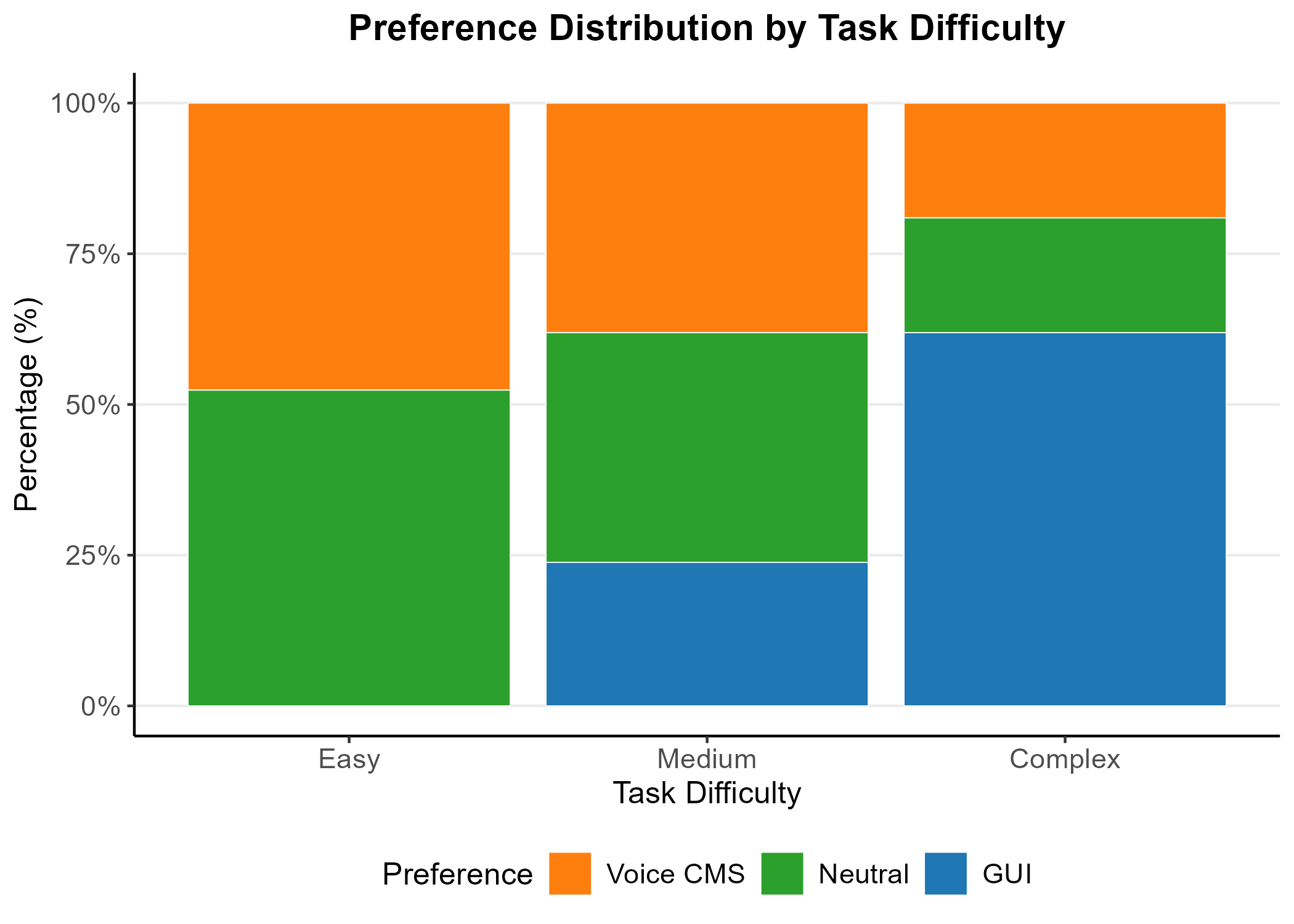}
   \caption{Preference Distribution by Task Difficulty}
   \label{fig:pref_difficulty}
\end{figure}

The initial hypothesis of the experiment was that user preference would shift towards the voice-controlled interface as task complexity increased. This assumption stemmed from the understanding that voice interfaces, with their natural language input, might reduce cognitive load during complex interactions. However, a preliminary analysis of preference distributions across task difficulty levels, visualized in Figure~\ref{fig:pref_difficulty}, suggested a potential trend reversal, with users possibly favoring the GUI for more complex tasks. It's important to acknowledge that this initial observation did not account for the repeated-measures design of the study, where each participant interacted with both interfaces across all difficulty levels. To examine the relationship between task complexity and interface preference while accounting for the nested structure of the data (126 task observations nested within 7 participants), a Bayesian multilevel categorical logistic regression model using the \texttt{brms} package in \texttt{R} was employed. This approach allows for isolating the effect of task difficulty on interface preference from individual variations in baseline preferences and testing the hypothesis that GUI is chosen more frequently as task complexity increases.

The model predicted the user's interface preference (a three-level categorical outcome: 'Neutral', 'GUI', 'Voice CMS', with 'Voice CMS' serving as the reference category) based on task difficulty (a three-level categorical predictor: 'Easy' as baseline, 'Medium', 'Complex'). Random intercepts were included for each participant (\texttt{scenarioId}) to account for individual differences in baseline preference patterns. The model was fitted using the No-U-Turn Sampler (NUTS) across 4 chains, and convergence diagnostics confirmed the reliability of the obtained estimates (all $\texttt{Rhat}\le 1.01$).

Significant variability in baseline preferences was observed across participants, confirming the necessity of the multilevel approach. The estimated standard deviation for the random intercepts was substantial both for the log-odds of preferring 'Neutral' versus 'Voice CMS' (SD = 2.41, 95\% CrI [1.08, 4.65]) and for preferring 'GUI' versus 'Voice CMS' (SD = 2.19, 95\% CrI [0.81, 4.70]). Examining the fixed effects revealed the influence of task difficulty on interface preference at the population level. For the baseline 'Easy' difficulty level, users showed a strong and statistically credible preference for 'Voice CMS' over 'GUI' (Intercept Log-Odds = -11.92, 95\% CrI [-36.27, -3.26]). The credible interval being entirely below zero indicates lower odds of choosing GUI compared to Voice CMS for easy tasks. No significant difference was found between 'Neutral' and 'Voice CMS' preference at this baseline difficulty (Intercept Log-Odds = 0.16, 95\% CrI [-2.02, 2.10]).

A key finding is that task difficulty significantly modulated the interface preference. Compared to 'Easy' tasks, the log-odds of preferring 'GUI' over 'Voice CMS' increased substantially and credibly for both 'Medium' difficulty tasks (Estimate = 11.21, 95\% CrI [2.62, 35.58]) and 'Complex' difficulty tasks (Estimate = 13.69, 95\% CrI [4.90, 38.23]). This indicates a strong shift towards preferring the GUI relative to Voice CMS as tasks become more complex. In contrast, the relative preference between 'Neutral' and 'Voice CMS' did not change significantly with increasing task difficulty (Medium vs. Easy: Log-Odds Change = -0.53, 95\% CrI [-1.71, 0.60]; Complex vs. Easy: Log-Odds Change = -0.88, 95\% CrI [-2.38, 0.49]).

In summary, while Voice CMS held a clear preference advantage for easy tasks, this advantage was overcome as task difficulty increased, with users becoming significantly more likely to prefer the GUI for medium and complex tasks relative to Voice CMS. The significant positive coefficients for the GUI preference at higher difficulty levels provide strong statistical support for the hypothesis that the GUI interface is increasingly preferred as tasks become more complex within this experimental context.

\subsection{Influence of SEQ Difference on Preference}

\begin{figure}
   \centering
   \includegraphics[width=0.6\linewidth]{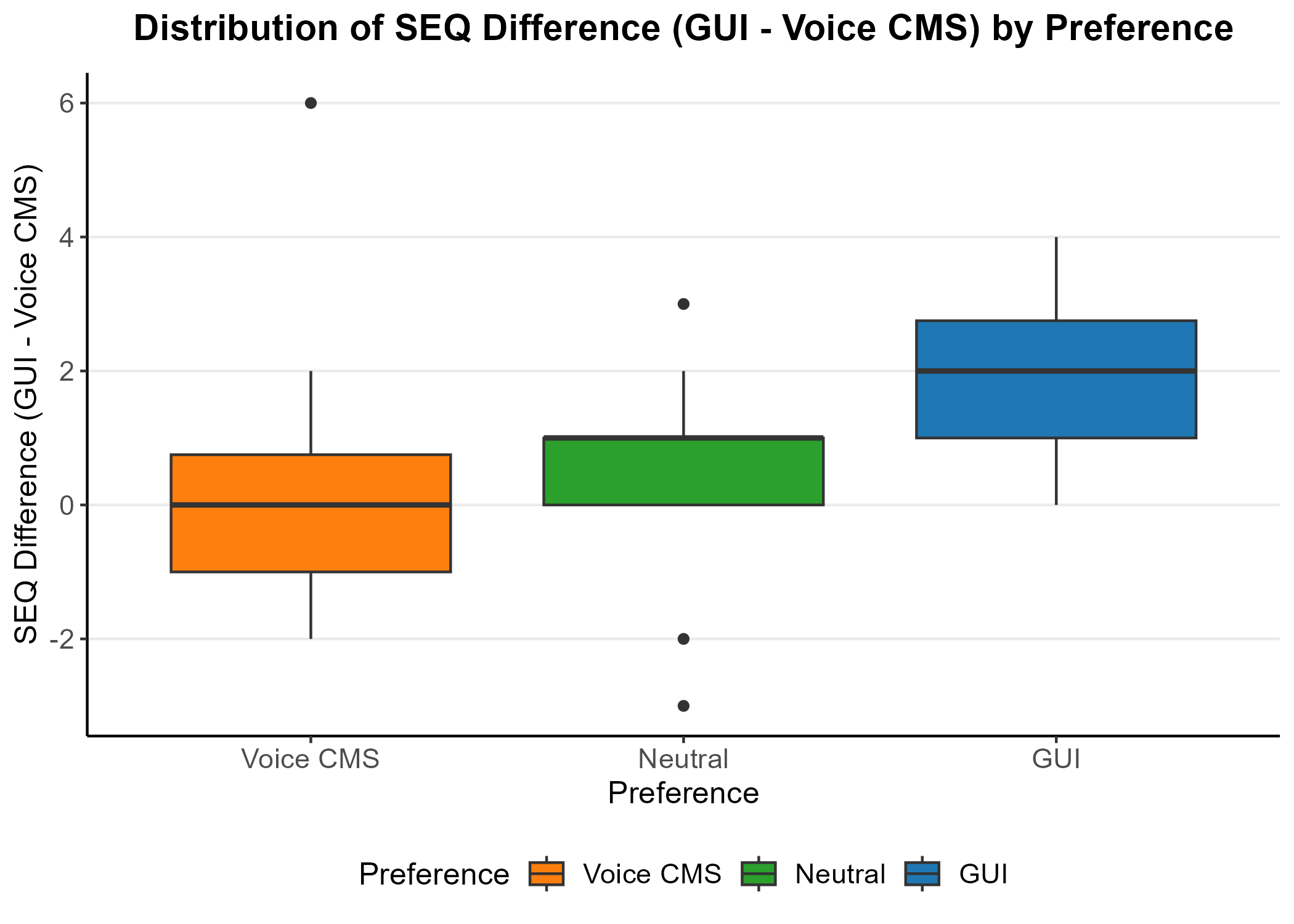}
   \caption{SEQ Difference (GUI - Voice CMS) by Preference}
   \label{fig:seqdiff_boxplot}
\end{figure}

To investigate how the perceived difference in task difficulty between the interfaces influenced user preference, we analyzed the relationship between \texttt{seq\_diff} (SEQ Score GUI - SEQ Score Voice CMS) and the chosen interface (\texttt{preference}). Initial exploration using visualizations (Figure~\ref{fig:seqdiff_boxplot}) suggested that Voice CMS or Neutral preferences were more common when the subjective difficulty was perceived as similar for both interfaces (low absolute \texttt{seq\_diff}), while GUI preference appeared to increase notably when it was rated as substantially easier than Voice CMS (positive \texttt{seq\_diff}). To formally test this relationship while accounting for non-independent observations from the 7 participants (\texttt{scenarioId}), two separate generalized linear mixed-effects models (GLMMs) with a binomial error distribution and logit link function were fitted using the \texttt{lme4} package in \texttt{R}. Both models included \texttt{seq\_diff} as a fixed effect predictor and a random intercept for \texttt{scenarioId}. The first model predicted the log-odds of preferring GUI versus other options (Neutral or Voice CMS), while the second predicted the log-odds of preferring Voice CMS versus other options (Neutral or GUI), based on N=63 observations where both scores were available.

The model predicting GUI preference revealed a significant effect of the subjective difficulty difference. When the interfaces were rated similarly difficult (\texttt{seq\_diff} = 0), users were significantly less likely to prefer the GUI (Intercept Estimate = -2.25, SE = 0.72, z = -3.13, p = 0.0018). However, there was a strong positive association between \texttt{seq\_diff} and choosing the GUI (Estimate = 1.15, SE = 0.41, z = 2.78, p = 0.0054). This indicates that as the GUI was perceived as relatively easier (i.e., \texttt{seq\_diff} increased), the odds of preferring the GUI increased - the corresponding odds ratio (OR) was 3.16 (95\% CI [1.40, 7.09]). Complementary results were found for the model predicting Voice CMS preference. The baseline preference for Voice CMS when perceived difficulty was equal (\texttt{seq\_diff} = 0) was not significantly different from preferring other options (Intercept Estimate = -0.28, SE = 0.34, z = -0.82, p = 0.411). There was a significant negative association between \texttt{seq\_diff} and choosing Voice CMS (Estimate = -0.61, SE = 0.23, z = -2.61, p = 0.0091). As the GUI was perceived as relatively easier (\texttt{seq\_diff} increased), the odds of preferring the Voice CMS significantly decreased with the corresponding odds ratio equal to 0.54 (95\% CI [0.34, 0.86]).

\begin{figure}
   \centering
   \includegraphics[width=0.6\linewidth]{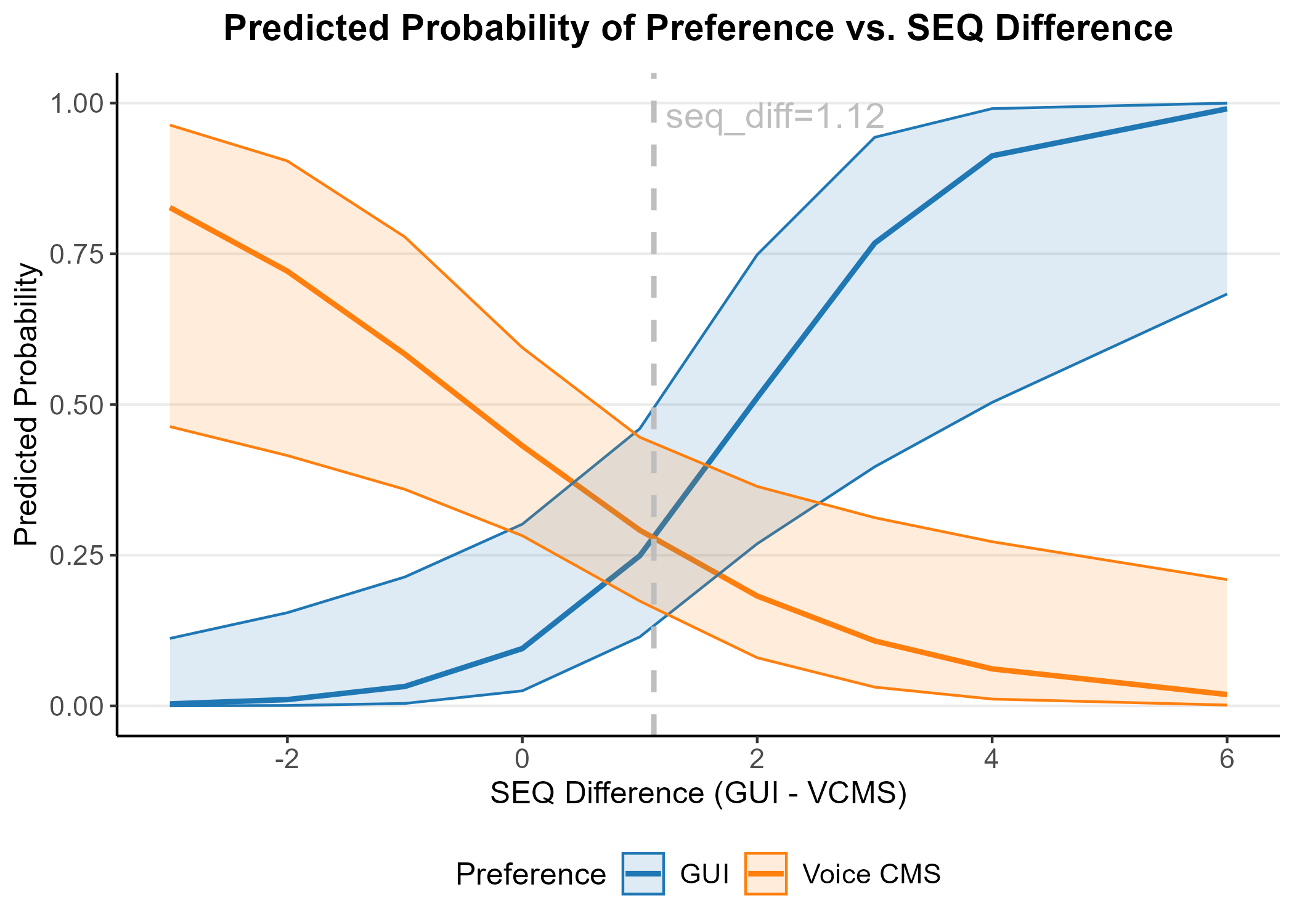}
   \caption{Predicted Probability of Preference vs SEQ Difference}
   \label{fig:seqdiff_combined}
\end{figure}

Both models indicated some variability between participants in their baseline preferences, as evidenced by the random intercept variances (Variance = 0.61 for GUI model, 0.11 for VCMS model). The predicted probability plot (Figure~\ref{fig:seqdiff_combined}) derived from these models further illustrates the findings. It shows that for tasks where subjective difficulty ratings were similar (\texttt{seq\_diff} near 0), the predicted probability of choosing Voice CMS was higher than that for GUI. The crossover point, where the predicted probability of choosing GUI surpasses that of Voice CMS, occurred when the SEQ score was just above 1.

In conclusion, the subjective difficulty difference (\texttt{seq\_diff}) was a significant predictor of preference, confirming that users generally tended to favour the interface they perceived as easier for a given task, even after accounting for user variability. This relationship, however, displayed an interesting asymmetry: Voice CMS held an advantage when perceived difficulty was comparable (GUI was less preffered at \texttt{seq\_diff}$\approx$0). Consequently, the GUI required a clear perceived ease-of-use advantage, specifically being rated more than $\approx$1 point easier on the 7-point SEQ scale, before it consistently surpassed the Voice CMS in user preference according to model predictions.

\subsection{Influence of Processing Time Difference on Interface Preference}

\begin{figure}
   \centering
   \includegraphics[width=0.6\linewidth]{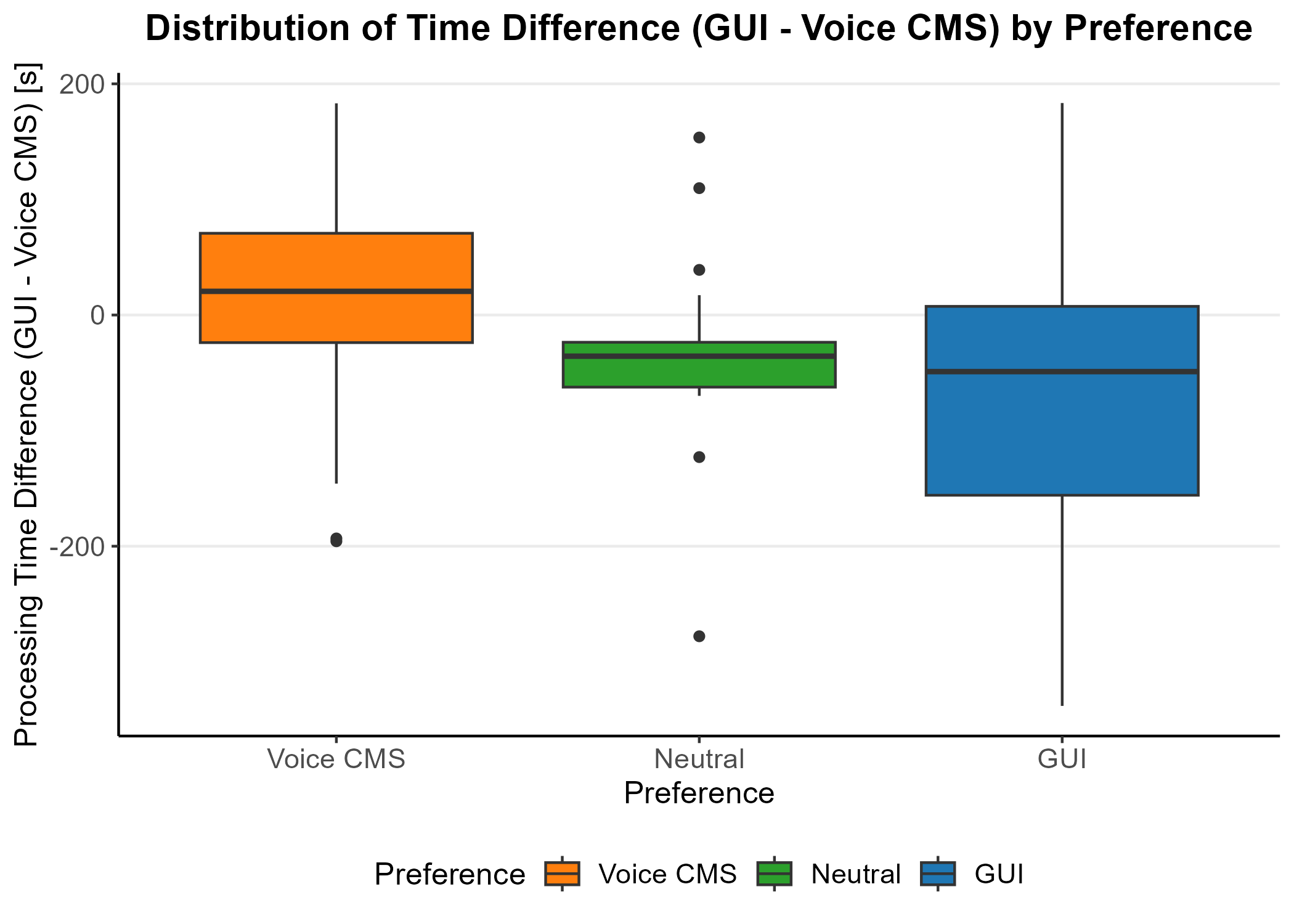}
   \caption{Processing Time Difference (GUI - Voice CMS) by Preference}
   \label{fig:timediff_boxplot}
\end{figure}

To further explore factors influencing user choice, the relationship between the difference in task completion time (\texttt{time\_diff} = GUI Time - Voice CMS Time) and interface preference (\texttt{preference}) was examined. Analysis of the distribution of completion time depending on preference (Figure~\ref{fig:timediff_boxplot}) suggests potential differences in how time influenced the choice between interfaces. To test the hypothesis that faster task completion favours an interface's preference, while accounting for user variability (\texttt{scenarioId}), two binomial generalized linear mixed-effects models (GLMMs) were fitted (N=63 observations, 7 users), predicting the preference for GUI and Voice CMS respectively, using \texttt{time\_diff} as the predictor.

Analysis of the random effects indicated moderate variability between participants in their baseline tendency to prefer the GUI interface (Intercept SD = 0.78), but considerably less variability in their baseline tendency to prefer the Voice CMS interface (Intercept SD = 0.24) after accounting for the time difference.

Examining the fixed effects, the model for GUI preference showed that when processing times were equal (\texttt{time\_diff} = 0), users were significantly less likely to prefer the GUI (Intercept Estimate = -1.23, SE = 0.47, z = -2.62, p = 0.009). A negative trend was observed for the \texttt{time\_diff} coefficient (Estimate = -0.0060, SE = 0.0033, z = -1.81, p = 0.071), indicating that the odds of preferring the GUI decreased as its relative completion time increased. This trend approached statistical significance (with OR = 0.994).

Conversely, the model for Voice CMS preference revealed a statistically significant positive effect of \texttt{time\_diff} (Estimate = 0.0059, SE = 0.0029, z = 2.08, p = 0.038). As the GUI took relatively longer (positive \texttt{time\_diff}), the odds of preferring Voice CMS significantly increased (OR = 1.006). The baseline preference for Voice CMS when times were equal showed only a non-significant tendency to be disfavoured (Intercept Estimate = -0.55, SE = 0.30, z = -1.85, p = 0.064).

\begin{figure}
   \centering
   \includegraphics[width=0.6\linewidth]{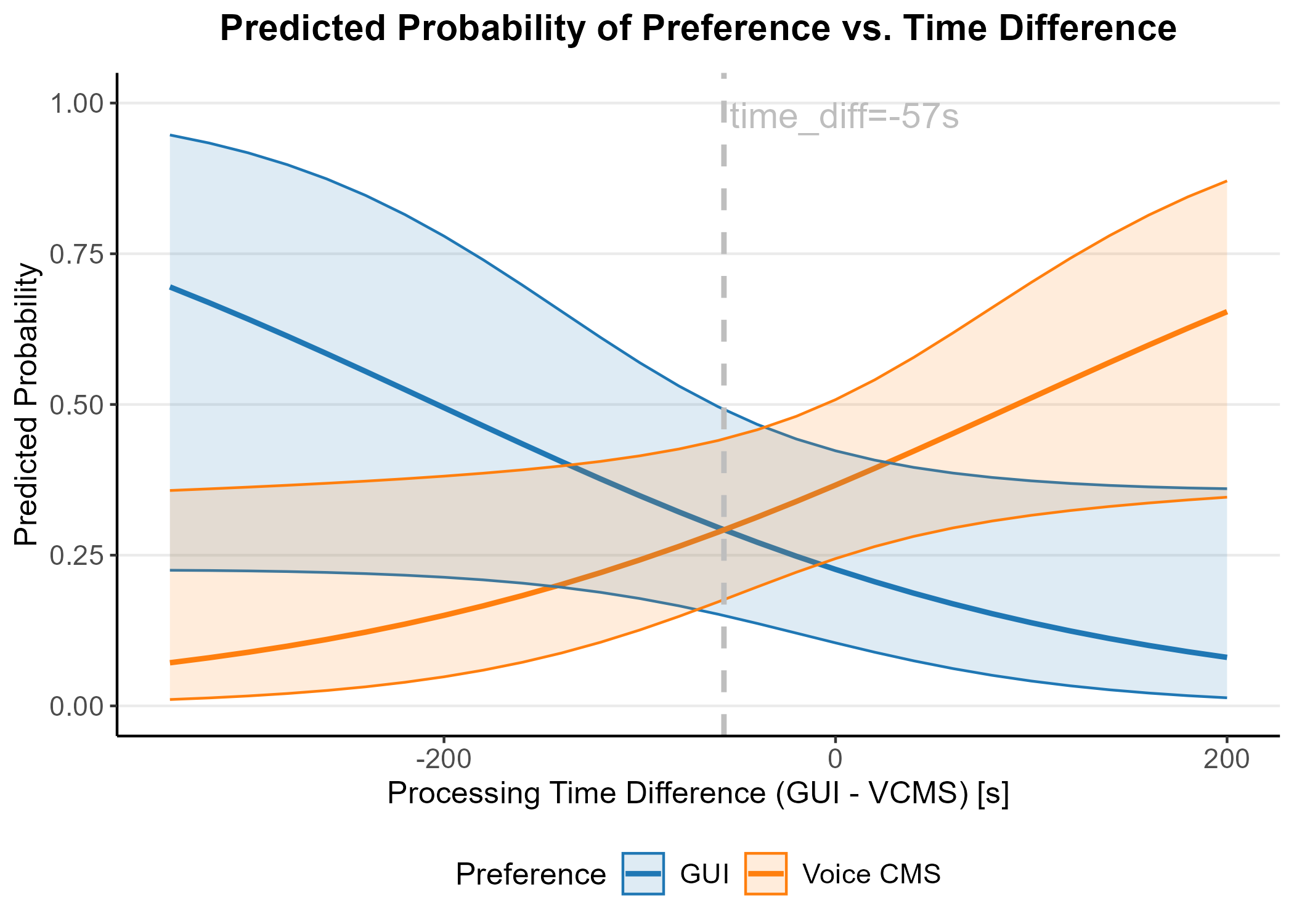}
   \caption{Predicted Probability of Preference vs Processing Time Difference}
   \label{fig:timediff_combined}
\end{figure}

Visualization of the predicted probabilities from these models in Figure~\ref{fig:timediff_combined} revealed an interesting crossover point. The predicted probability of choosing the GUI only surpassed that of choosing the Voice CMS when the completion time in the GUI was noticeably faster. This intersection occurred at a time difference of approximately -57 seconds.

In conclusion, the analysis indicates that while preference generally shifted towards the faster interface, the effect was statistically robust for favouring Voice CMS when the GUI was slower (p = 0.038) and showed a strong trend towards favouring the GUI when it was faster (p = 0.071). This dynamic, supporting the "faster is better" hypothesis, must be viewed alongside the baseline finding: the GUI was significantly less preferred at equal speeds and required completing tasks approximately 57 seconds faster to become the more probable choice.

\subsection{Voice CMS Interaction Metrics}

This subsection details the analysis of user interaction patterns specifically with the Voice CMS interface, focusing on the number of messages exchanged between experiment participants and the digital assistant, the number of summaries required before accepting the correctness of gathered information, and the influence of task characteristics like difficulty and length. The analysis is based on the 63 observations collected during Voice CMS interactions.

Statistics from the experiment data revealed the nature of user interactions with the Voice CMS. The number of summaries (at least one for each task) per single task was relatively low, with an average of 1.43 (SD = 1.00, Median = 1, Range [1, 7]). Notably, the analysis showed that partial summaries, a part of a user feedback mechanism after each new part of information received by the assistant, were consistently enabled during message exchanges in this dataset. The average total number of messages exchanged per task was 4.3 (SD = 2.76, Median = 4, Range [2, 16]), and it was identical to the number of messages sent with summaries on.

Both predefined task difficulty and the objective task length (measured in characters) were found to be strongly related, with task length increasing significantly across 'Easy' (M = 94.7), 'Medium' (M = 284.0), and 'Complex' (M = 425.0) categories (LMM fixed effects for difficulty: p < 0.01).

Task difficulty significantly influenced interaction volume. The mean number of total messages increased progressively from 'Easy' (M = 2.76) through 'Medium' (M = 4.19) to 'Complex' (M = 5.95) tasks. A Generalized Linear Mixed Model (GLMM) confirmed that difficulty level was a significant positive predictor of the number of messages exchanged (Poisson GLMM, linear effect of difficulty: Estimate = 0.54, p < 0.001), after accounting for random intercepts per user. A similar, though less pronounced, increasing trend was observed for the mean number of summaries generated across difficulty levels (Easy=1.05, Medium=1.48, Complex=1.76).

Furthermore, task length independently predicted interaction metrics, even after controlling for the categorical difficulty level and user variability. Longer tasks were significantly associated with longer processing times (LMM, Estimate = 0.87 s/char, p < 0.001) and a higher number of total messages exchanged (Poisson GLMM, Estimate = 0.0027 log-count/char, p = 0.033). 

\begin{figure}
   \centering
   \includegraphics[width=0.6\linewidth]{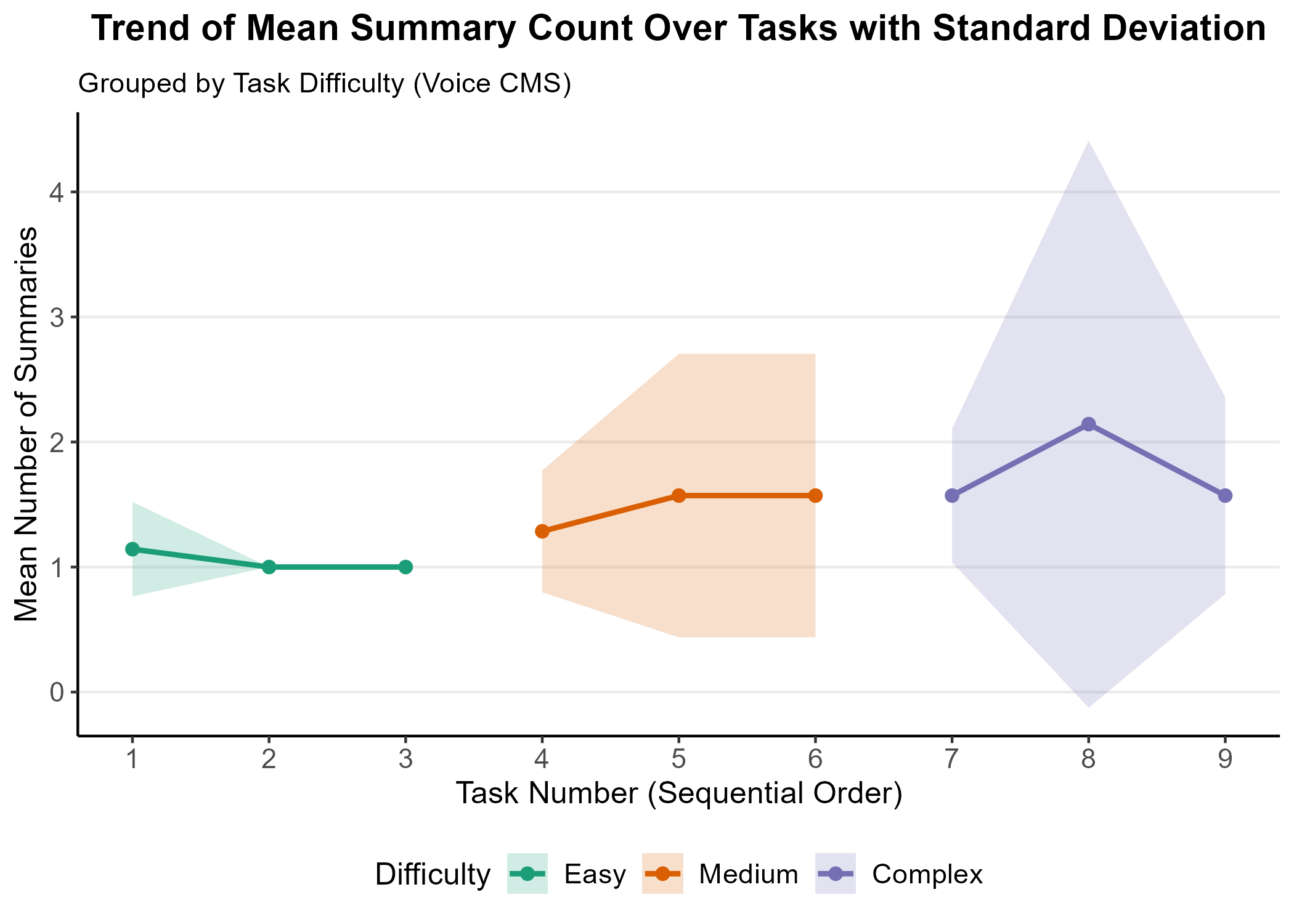}
   \caption{Trend of Mean Summary Count over Tasks}
   \label{fig:summary_trend}
\end{figure}

The tasks were executed in a random (but fixed order). The plot in Figure~\ref{fig:summary_trend} displays them grouped by difficulty but without changing relative order. Visual inspection of interaction metrics did not reveal consistent learning effects or clear trends indicating systematic changes in message (plot not provided) or summary counts as users progressed through the experiment. The study was not designed in a way that allowed for measuring the impact of the learning interface and gaining experience.

Regarding performance outcomes, the number of summaries generated was not found to be a significant predictor of the objective task score when controlling for task difficulty and user effects in a Linear Mixed Model (LMM, p = 0.97). Finally, confirming the intuitive relationship between interaction volume and duration, the total number of messages exchanged was a strong, significant positive predictor of task completion time (LMM, Estimate = 40.6 s/message, p < 0.001), even when accounting for task difficulty and user variability.

In summary, interactions with the Voice CMS were strongly influenced by task characteristics. Both increasing task difficulty and objective task length led to significantly longer interactions (more messages) and increased processing times. Task difficulty was linked to the number of required summaries, but this number did not significantly impact task score in this analysis.

\section{Discussion}
%proven new knowledge, limitations, porównanie do istniejących badań SUS na czymś porównywalnym (dodałem referencje do Zotero)

The comparative analysis revealed distinct performance and usability profiles for the GUI and Voice CMS in the context of knowledge management tasks. Although objective accuracy was similar, some differences emerged in perceived usability, task completion speed, and notably, in user preference patterns. These differences were not uniform but were influenced by the complexity of the tasks performed, prompting a closer look at how task demands interacted with interface choice.

%\paragraph{Revisiting the Initial Hypothesis: The Role of Task Complexity}
Our central hypothesis posited that users would prefer the Voice CMS over the GUI for more complex tasks, assuming the natural language interaction would alleviate the cognitive load associated with structuring complex information typical in GUI forms. However, our statistical analysis contradicted this initial assumption. The multilevel categorical logistic regression model showed that while Voice CMS was indeed preferred over GUI for 'Easy' tasks (where GUI was never chosen as the preferred option), preference significantly shifted towards the GUI as task difficulty increased from 'Easy' to 'Medium' and 'Complex'. This suggests that, within the context of this study, factors potentially favouring the GUI --- such as visual oversight, direct manipulation capabilities for precise data entry (dates, specific entities mentioned by users), ease of error correction, or potentially greater user familiarity with graphical paradigms --- outweighed the hypothesised benefits of conversational input for managing increased task complexity. User comments align with this, noting difficulties in conveying complex, nuanced information via voice and challenges in correcting ASR/NLU misinterpretations (\textit{"conveying intricacies... can be difficult and time-consuming"}, \textit{"hard to correct afterwards"}).

%\paragraph{Subjective Perceptions, Efficiency, and Preference Thresholds}
Despite the overall trend favouring GUI for harder tasks, the analyses of subjective ease-of-use difference and processing time difference revealed a more nuanced picture of user preference. While users generally preferred the interface they perceived as easier or faster, the Voice CMS demonstrated a degree of resilience. The analysis showed that the GUI needed to be perceived as more than one point easier on the 7-point SEQ scale before it became the statistically more probable choice. Similarly, the GUI had to be substantially faster, completing the task approximately 57 seconds quicker than the Voice CMS, to overcome a baseline disadvantage and surpass the voice interface in predicted preference probability. This suggests that users might tolerate a degree of perceived inefficiency or difficulty with the Voice CMS. This tolerance could stem from the perceived benefit of reduced physical typing effort, the novelty, or the more natural interaction, as hinted by comments describing the voice interface as an \textit{"attractive alternative"} that \textit{"reduces the effort associated with typing"}. A deeper investigation into the factors driving this tolerance would be a valuable direction for future work to fully understand the adoption potential of Voice CMS.

%\paragraph{Voice Interface Interaction Dynamics: The Critical Role of Feedback}
Analysis specific to the Voice CMS interactions highlighted the importance of feedback and the impact of task characteristics on dialogue patterns. The number of messages exchanged and overall processing time increased with both predefined task difficulty and objective task length, confirming that more complex or longer tasks required more extensive interaction. Notably, users consistently utilised the partial summary feature provided after each piece of information was processed by the assistant. This universal usage underscores the perceived necessity of feedback for verification and control in a voice-only interaction, a point strongly echoed in user comments (\textit{"wanted to be sure I entered it correctly"}, \textit{"wouldn't turn off summary, not because of AI itself"}). However, while essential for user confidence, the number of summaries generated was not significantly associated with objective task success, suggesting that feedback alone did not guarantee perfect outcomes, perhaps due to difficulties in catching subtle errors (\textit{"harder to catch errors like 8.20 instead of 8-20"}) or correcting them effectively via voice. 

%\paragraph{Limitations and Future Directions}
Several limitations should be considered when interpreting these findings. The study involved a small sample size (N=7), limiting the generalizability of the results. The interaction was short-term, preventing conclusions about long-term usability, learning effects, or adaptation to the Voice CMS, which users commented might improve their experience (\textit{"would require 'training'"}, \textit{"if I knew... my rating would be higher"}). Although tasks were presented textually, the inherent structure might have unintentionally favoured GUI interaction patterns if users mentally mapped them to form-filling, potentially underutilizing the conversational capabilities of the Voice CMS, as one user speculated (\textit{"impression might change if information wasn't provided so concisely"}). Furthermore, the Voice CMS prototype exhibited technical limitations noted by users (ASR/NLU errors, interruptions), which likely impacted usability scores and preferences. Finally, while not measured, participants likely had significantly more experience with GUIs, potentially influencing their baseline preferences and performance. Future research should involve a larger participant pool, incorporate improvements addressing the reported limitations, and examine long-term adoption dynamics and learning curves associated with Voice CMS use.

Despite these limitations, the study highlights the potential of voice interfaces as an "attractive alternative" for data entry tasks, particularly simpler ones --- this attractiveness becomes even more relevant in a hotel setting, where GUI deployment could be less feasible. The strong user reliance on summaries, coupled with comments about the burden of auditory verification and the desire for visual confirmation, points towards a promising avenue for future research and development: hybrid interfaces. As suggested by multiple participants, combining voice input for its naturalness and efficiency in capturing information with a synchronized visual display for real-time verification and summary could mitigate the core weaknesses observed in the voice-only modality. Such a hybrid system might significantly reduce interaction time (by lessening the need for lengthy spoken summaries) and enhance user confidence, potentially shifting preference towards voice-centric workflows even for more complex tasks. Future research should also explore these hybrid designs.

\section{Conclusions}
This paper presents a nuanced view of user preference between GUI and Voice CMS interfaces. While the GUI held advantages in overall usability ratings, subjective ease-of-use, and efficiency, particularly as task difficulty increased, the Voice CMS demonstrated strong user preference for simpler tasks and a degree of user tolerance for moderate inefficiencies. The need for robust feedback in voice interactions was evident. The results suggest that while voice interfaces hold significant potential as an alternative for data entry, overcoming challenges related to handling complexity and ensuring user confidence through effective, potentially visual, feedback mechanisms is key to broader adoption. Hybrid voice-visual interfaces represent a compelling direction to harness the strengths of conversational input while providing the assurance that users desire in the visual form, and appear to be a particularly promising direction for future development and research.

\bibliographystyle{unsrtnat}
\bibliography{references}  %%% Uncomment this line and comment out the ``thebibliography'' section below to use the external .bib file (using bibtex) .





% \end{thebibliography}

\end{document}